\documentclass[a4paper, 10pt, conference]{ieeeconf}      
\usepackage{amssymb,amsmath,epsfig,graphicx,theorem,threeparttable}
\usepackage{amsfonts}
\usepackage{bm}
\usepackage{cite}

\newcommand{\pv}{\mathbf{p}}

\IEEEoverridecommandlockouts                              
\title{\LARGE \bf
The Capacity Region of a Class of Discrete
Degraded Interference Channels
}

\author{Nan Liu and Sennur Ulukus
\thanks{This work was supported by
NSF Grants CCR $03$-$11311$, CCF $04$-$47613$ and CCF
$05$-$14846$.}
\thanks{N. Liu and S. Ulukus are with the Department of Electrical and Computer Engineering,
        University of Maryland, College Park, MD 20742, USA
        {\tt\small nkancy@umd.edu; ulukus@umd.edu}}%
}

\begin{document}

\maketitle
\thispagestyle{empty}
\pagestyle{empty}

\begin{abstract}

We provide a single-letter characterization for the capacity
region of a class of discrete degraded interference channels
(DDICs). The class of DDICs considered includes the discrete
additive degraded interference channel (DADIC) studied by
Benzel\cite{Benzel:1979}. We show that for the class of DDICs
studied, encoder cooperation does not increase the capacity
region, and therefore, the capacity region of the class of DDICs
is the same as the capacity region of the corresponding degraded
broadcast channel.
\end{abstract}

\section{Introduction}
In wireless communications, where multiple transmitter and
receiver pairs share the same medium, interference is unavoidable.
How to best manage interference coming from other users
 and how not to cause too much interference to other users while maintaining the quality of communication
 is a challenging question and of a great deal of practical interest.

To be able to understand the effect of interference on
communications better, interference channel (IC) has been
introduced in \cite{Shannon:1961}. The IC is a simple network
consisting of two pairs of transmitters and receivers. Each pair
wishes to communicate at a certain rate with negligible
probability of error. However, the two communications interfere
with each other. To best understand the management of
interference, we need to find the capacity region of the IC.
However, the problem of finding the capacity region of the IC is
essentially open except in some special cases, e.g., a class of
deterministic ICs \cite{ElGamal:1982}, discrete additive degraded
interference channels (DADICs) \cite{Benzel:1979}, strong ICs
\cite{Sato:1981,Costa:1987}, ICs with statistically equivalent
outputs \cite{Carleial:1978,Sato:1977,Ahlswede:1971}.

In this paper, we consider a class of
discrete degraded interference channels (DDICs). In a DDIC, only
the ``bad'' receiver faces interference, while the ``good''
receiver has the ability to decode both messages and thus, behaves
like the receiver of a multiple access channel. It is this fact
that makes the DDIC easier to analyze as compared to the IC, where
both receivers are faced with interference.

We provide a single-letter characterization for the capacity
region of a class of DDICs. The class of DDICs includes the DADICs
studied by Benzel\cite{Benzel:1979}. We show that for the class of
DDICs studied, encoder cooperation does not increase the capacity
region, and therefore, the capacity region of the class of DDICs
is the same as the capacity region of the corresponding degraded
broadcast channel, which is known.

\section{System Model}
A discrete memoryless IC consists of two transmitters and two
receivers. Transmitter $1$ has message $W_1$ to send to receiver
$1$. Transmitter $2$ has message $W_2$ to send to receiver $2$.
Messages $W_1$ and $W_2$ are independent. The channel consists of
two input alphabets, $\mathcal{X}_1$ and $\mathcal{X}_2$, and two
output alphabets, $\mathcal{Y}_1$ and $\mathcal{Y}_2$. The channel
transition probability is $p(y_1,y_2|x_1,x_2)$.

In this paper, our definition of degradedness is in the stochastic sense, i.e., we say that
an IC is DDIC if there exists a probability
distribution $p'(y_2|y_1)$ such that
\begin{align}
p(y_2|x_1,x_2)=\sum_{y_1 \in \mathcal{Y}_1} p(y_1|x_1,x_2)
p'(y_2|y_1) \label{realsto}
\end{align}
for all $x_1 \in \mathcal{X}_1$, $x_2 \in \mathcal{X}_2$ and $y_2
\in \mathcal{Y}_2$.
However, we note that for any DDIC, we can form another DDIC (physically degraded) by
\begin{align}
p(y_1,y_2|x_1,x_2)=p(y_1|x_1,x_2)p'(y_2|y_1) \label{sto}
\end{align}
which has the same marginals, $p(y_1|x_1,x_2)$
and $p(y_2|x_1,x_2)$, as the original DDIC.
Since the receivers do not cooperate in an IC, similar
to the case of the broadcast channel \cite[Problem
14.10]{Cover:book}, the capacity region is only a function of the
marginals, $p(y_1|x_1,x_2)$ and $p(y_2|x_1,x_2)$, and the rate pairs
in the capacity region can be achieved by the same achievability
scheme for different ICs with the same marginals. Hence, the capacity results that we obtain for DDICs which
satisfy (\ref{sto}) will be valid for any DDIC that has the
same marginals, $p(y_1|x_1,x_2)$ and $p(y_2|x_1,x_2)$. Thus, without loss of generality, from now on, we may restrict ourselves to studying
DDICs that satisfy (\ref{sto}).

A DDIC is characterized by two transition
probabilities, $p'(y_2|y_1)$ and $p(y_1|x_1,x_2)$. For notational
convenience, let $T'$ denote the $|\mathcal{Y}_2| \times
|\mathcal{Y}_1|$ matrix of transition probabilities $p'(y_2|y_1)$, and
$T_{\bar{x}_2}$ denote the $|\mathcal{Y}_1| \times
|\mathcal{X}_1|$ matrix of transition probabilities
$p(y_1|x_1,\bar{x}_2)$, for all $\bar{x}_2 \in \mathcal{X}_2$.

Throughout the paper, $\Delta_n$ will denote the probability simplex
\begin{align}
\bigg\{(p_1,p_2, \cdots, p_n)\bigg|\sum_{i=1}^n p_i=1, \quad p_i
\geq 0,
 i=1,2,\cdots,n  \bigg\}
\end{align}
and $\mathcal{J}_n$ will denote the representation of the symmetric
group of permutations of $n$ objects by the $n \times n$
permutation matrices.

The class of DDICs we consider in this paper satisfies the
following conditions:
\begin{enumerate}
\item $T'$ is input symmetric. Let the input symmetry group be
$\mathcal{G}$. \item For any $x_2', x_2'' \in \mathcal{X}_2$,
there exists a permutation matrix $G \in \mathcal{G}$, such that
\begin{align}
T_{x_2'}=G T_{x_2''}
\end{align}
\item $H(Y_1|X_1=x_1, X_2=x_2)=\eta$, independent of $x_1$, $x_2$.
\item $p(y_1|x_1,x_2)$ satisfies
\begin{align}
\sum_{x_2} p(y_1|x_1,x_2)=\frac{|\mathcal{X}_2|}{|\mathcal{Y}_1|},
\quad x_1 \in \mathcal{X}_1, y_1 \in \mathcal{Y}_1
\label{condition41}
\end{align}
\item Let $\pv_{x_1,x_2}$ be the $|\mathcal{Y}_1|$ dimensional
vector of probabilities $p(y_1|x_1,x_2)$ for a given $x_1,x_2$.
Then, there exists an $\tilde{x}_2 \in \mathcal{X}_2$, such that
\begin{align}
&\bigg\{\sum_{x_1,x_2} a_{x_1,x_2} \pv_{x_1,x_2}: \sum_{x_1,x_2}
a_{x_1,x_2}=1, a_{x_1,x_2} \geq 0\bigg\} \nonumber \\
\subseteq &\bigg\{G \bigg( \sum_{x_1} b_{x_1}
\pv_{x_1,\tilde{x}_2}\bigg): \sum_{x_1} b_{x_1}=1, b_{x_1} \geq 0,
G \in \mathcal{G} \bigg\} \label{condition42}
 \end{align}
\end{enumerate}

The definition of an input symmetric channel is given in
\cite[Section II.D]{Witsenhausen:1975}. For completeness, we
repeat it here. For an $m \times n$ stochastic matrix $T'$ (an $n$
input, $m$ output channel), the input symmetry group $\mathcal{G}$
is defined as
\begin{align}
\mathcal{G}= \{G \in \mathcal{J}_n: \exists \Pi \in \mathcal{J}_m,
\quad T'G=\Pi T'\}
\end{align}
i.e., $\mathcal{G}$ is the set of permutation matrices $G$ such that the
column permutations of $T'$ with $G$ may be achieved with
corresponding row permutations. $T'$ is input symmetric, if
$\mathcal{G}$ is transitive, i.e., any element of
$\{1,2,\cdots,n\}$ can be mapped to every other element of
$\{1,2,\cdots,n\}$ by some member of $\mathcal{G}$. $\mathcal{G}$
being a transitive subgroup means that the output entropy of
channel $T'$ is maximized when the input distribution is chosen to
be the uniform distribution, i.e.,
\begin{align}
\max_{\pv \in \Delta_n} H(T'\pv)=H(T'\mathbf{u}) \label{houjia}
\end{align}
where $\mathbf{u}$ denotes the uniform distribution in $\Delta_n$.
This is because, for any $\pv \in \Delta_n$, if we let
$\mathbf{q}=|\mathcal{G}|^{-1} \sum_{G \in \mathcal{G}} G \pv$,
then we have
\begin{align}
H(T' \mathbf{q}) &= H \left( |\mathcal{G}|^{-1}
\sum_{G \in \mathcal{G}} T'G \pv\right) \\
&= H \left( |\mathcal{G}|^{-1}
\sum_{G \in \mathcal{G}} \Pi_{G} T' \pv\right)  \label{space1}\\
& \geq |\mathcal{G}|^{-1}
\sum_{G \in \mathcal{G}} H \left(\Pi_{G} T' \pv \right) \label{space2}\\
&=H(T' \pv)
\end{align}
where (\ref{space1}) follows from the fact that $G \in
\mathcal{G}$, and (\ref{space2}) follows from the concavity of the
entropy function. Note that for any $G' \in \mathcal{G}$,
\begin{align}
G' \mathbf{q}=\mathbf{q}
\end{align}
by the fact that $\mathcal{G}$ is a group. Since $\mathcal{G}$ is
also transitive, $\mathbf{q}=\mathbf{u}$.

Condition 2 implies that for any $p(x_1)$, $ H(Y_1|X_2=x_2) $ does
not depend on $x_2$. Combined with condition 1, condition 2
further implies that $ H(Y_2|X_2=x_2) $ does not depend on $x_2$
either. These two facts will be proved and utilized in other
proofs later.

A sufficient condition for condition 3 to hold is that the vectors
$p(y_1|X_1=x_1,X_2=x_2)$ for all $(x_1,x_2) \in \mathcal{X}_1
\times \mathcal{X}_2$ are permutations of each other. This is true
for instance when the channel from $Y_1$ to $Y_2$ is additive
\cite{Benzel:1979}.

By condition 4, we can show that when $X_2$ takes the uniform
distribution, $Y_1$ will also be uniformly distributed. Combined
with condition 1, condition 4 implies that when $X_2$ takes the
uniform distribution, $H(Y_2)$ is maximized, irrespective of
$p(x_1)$.

In condition 5, the first line of (\ref{condition42}) denotes the
set of all convex combinations of vectors $\pv_{x_1,x_2}$ for all
$x_1,x_2 \in \mathcal{X}_1 \times \mathcal{X}_2$, while the second
line denotes all convex combinations, and their permutations with
$G \in \mathcal{G}$, of vectors $\pv_{x_1,\tilde{x}_2}$ for all
$x_1 \in \mathcal{X}_1$, but for a fixed $\tilde{x}_2 \in
\mathcal{X}_2$. Therefore, this condition means that all convex
combinations of $\pv_{x_1,x_2}$ may be obtained by a combination
of convex combinations of $\pv_{x_1,\tilde{x}_2}$ for a fixed
$\tilde{x}_2$, and permutations in $\mathcal{G}$.

The DADICs considered in \cite{Benzel:1979} satisfy conditions
1-5, as we will show in Section \ref{first}.

The aim of this paper is to provide a single-letter
characterization for the capacity region of DDICs that satisfy
conditions 1-5, and we will follow the proof technique of
\cite{Benzel:1979} with appropriate generalizations.

\section{The Outer Bound (Converse)}
When we assume that the encoders are able to fully cooperate,
i.e., both encoders know both messages $W_1$ and $W_2$, we get a
corresponding degraded broadcast channel with input $x=(x_1,x_2)$.
The capacity region of the corresponding degraded broadcast
channel serves as an outer bound on the capacity region of the
DDIC. The capacity region of the degraded broadcast channel is
known \cite{Cover:1972,Gallager:1965,Cover:book}, and thus, a
single-letter outer bound on the capacity region of the DDIC is
\begin{align}
\overline{\text{co}} \Bigg[ \bigcup_{p(u),p(x_1,x_2|u)}
\bigg\{(R_1,R_2): R_1 &\leq I(X_1, X_2;Y_1|U) \nonumber \\
 R_2 &\leq I(U;Y_2)   \bigg\}\Bigg] \label{outer1}
\end{align}
where $\overline{\text{co}}$ denotes the closure of the convex
hull operation, and the auxiliary random variable $U$, which
satisfies the Markov chain $U \longrightarrow (X_1,X_2)
\longrightarrow Y_1 \longrightarrow Y_2$, has cardinality bounded
by $|\mathcal{U}| \leq \min \left(|\mathcal{Y}_1|,|\mathcal{Y}_2|,
|\mathcal{X}_1| |\mathcal{X}_2| \right)$. More specifically, for
DDICs that satisfy condition 3, (\ref{outer1}) can be written as
\begin{align}
\overline{\text{co}} \Bigg[
 \bigcup_{  p(u),p(x_1,x_2|u) }\bigg\{ (R_1,R_2): R_1 &\leq H(Y_1|U)-\eta \nonumber \\
  R_2 &
\leq I(U;Y_2) \bigg\} \Bigg] \label{outer2}
\end{align}
Let us define $T(c)$ as
\begin{align}
T(c)=\max_{\begin{array}{c}p(u)p(x_1,x_2|u) \\
H(Y_1|U)=c  \\
|\mathcal{U}| \leq \min \left(|\mathcal{Y}_1|,|\mathcal{Y}_2|,
|\mathcal{X}_1|  |\mathcal{X}_2| \right)\end{array}}I(U;Y_2)
\label{defineT}
\end{align}
where the entropies are calculated according to the distribution
\begin{align}
p(u,x_1,x_2,y_1,y_2)=p(u)p(x_1,x_2|u)p(y_1|x_1,x_2)p'(y_2|y_1)
\end{align}
Using condition 3, we can show that $\eta \leq c \leq \log
|\mathcal{Y}_1|$. $T(c)$ is concave in $c$
\cite{Benzel:1979,Ahlswede:1975}, and therefore, (\ref{outer2})
can also be written as
\begin{align}
\bigcup_{\eta \leq c \leq \log |\mathcal{Y}_1|} \bigg\{(R_1,R_2):
R_1 &\leq c-\eta \nonumber \\
R_2 & \leq T(c)\bigg\} \label{outerreal}
\end{align}

\section{An Achievable Region}
Based on \cite[Theorem 4]{Sato:1977}, the following region is
achievable,
\begin{align}
\overline{\text{co}} \Bigg[ \bigcup_{p(x_1),p(x_2)}\bigg\{
(R_1,R_2): R_1 &\leq I(X_1;Y_1|X_2) \nonumber \\
 R_2 & \leq I(X_2;Y_2)
\bigg\}  \Bigg]\label{inner1}
\end{align}
which corresponds to the achievability scheme that the ``bad''
receiver treats the signal for the ``good'' receiver as pure
noise, and the ``good'' receiver decodes both messages as if it is
the receiver in a multiple access channel.

For DDICs that satisfy condition 3, (\ref{inner1}) reduces to
\begin{align}
\overline{\text{co}}\Bigg[ \bigcup_{p(x_1),p(x_2)} \bigg\{
(R_1,R_2): R_1 & \leq H(Y_1|X_2)-\eta \nonumber \\
 R_2 & \leq
H(Y_2)-H(Y_2|X_2) \bigg\} \Bigg] \label{inner2}
\end{align}
We note that (\ref{inner2}) remains an achievable region if we
choose $p(x_2)$ to be the uniform distribution. Furthermore, by
choosing $p(x_2)$ as the uniform distribution, we have
\begin{align}
p(y_1) = &\sum_{x_1,x_2} p(y_1|x_1,x_2) p(x_1)
\frac{1}{|\mathcal{X}_2|} \\
= &\frac{1}{|\mathcal{X}_2|} \sum_{x_1} p(x_1)
\sum_{x_2} p(y_1|x_1,x_2)  \\
 = & \frac{1}{|\mathcal{Y}_1|} \label{IEEE1}
\end{align}
where (\ref{IEEE1}) uses condition 4. Thus, when $p(x_2)$ is
chosen as the uniform distribution, $p(y_1)$ results in a uniform
distribution as well. Let us define $\tau$ as
\begin{align}
\tau= \max_{\pv \in \Delta_{|\mathcal{Y}_1|}} H(T'\pv)
\label{later11}
\end{align}
Using the fact that the DDIC under consideration satisfies
condition 1, i.e., it satisfies (\ref{houjia}), we have that when
$p(x_2)$ is uniform, and consequently $p(y_1)$ is uniform,
\begin{align}
H(Y_2)=\tau
\end{align}
Hence, choosing $p(x_2)$ to be the uniform distribution in
(\ref{inner2}), yields the following as an achievable region,
\begin{align}
\overline{\text{co}} \Bigg[\bigcup_{p(x_1)} \bigg\{(R_1,R_2): R_1
&\leq \frac{1}{|\mathcal{X}_2|}\sum_{x_2} H(Y_1|X_2=x_2)-\eta \nonumber \\
 R_2 &
\leq \tau-\frac{1}{|\mathcal{X}_2|}\sum_{x_2}H(Y_2|X_2=x_2)
\bigg\}\Bigg] \label{inner3}
\end{align}
Due to condition 2, for any $p(x_1)=\pv$ and any $x_2', x_2'' \in
\mathcal{X}_2$, there exists a permutation matrix $G \in
\mathcal{G}$ such that
\begin{align}
H(Y_1|X_2=x_2')& =H(T_{x_2'}\pv) \\
&=H(GT_{x_2''}\pv) \\
&=H(T_{x_2''}\pv) \\
&=H(Y_1|X_2=x_2'') \label{tildex2}
\end{align}
which means that for any $p(x_1)$, $H(Y_1|X_2=x_2)$ does not
depend on $x_2$. Furthermore, for any $p(x_1)=\pv$ and any $x_2',
x_2'' \in \mathcal{X}_2$, there exist permutation matrices $G \in
\mathcal{G}$ and $\Pi$, of order $|\mathcal{Y}_1|$ and
$|\mathcal{Y}_2|$ respectively, such that
\begin{align}
H(Y_2|X_2=x_2')& =H(T'T_{x_2'}\pv)\\
&=H(T'GT_{x_2''}\pv)\\
&= H(\Pi T' T_{x_2''}\pv) \label{IEEE2}\\
&=H(T'
T_{x_2''}\pv)\\
&=H(Y_2|X_2=x_2'') \label{usecon2}
\end{align}
where (\ref{IEEE2}) follows from the fact that $G \in \mathcal{G}$. (\ref{usecon2})
 means that for any $p(x_1)$, $H(Y_2|X_2=x_2)$ does not depend on $x_2$ either.
Hence, the achievable region in (\ref{inner3}) can further be
written as
\begin{align}
\overline{\text{co}} \Bigg[ \bigcup_{p(x_1)} \bigg\{ (R_1,R_2):
R_1 & \leq H(Y_1|X_2=x_2)-\eta \nonumber \\
R_2 & \leq \tau-H(Y_2|X_2=x_2) \bigg\} \Bigg] \label{innerwho}
\end{align}
for any $x_2 \in \mathcal{X}_2$. Since we will use condition 5
later, we choose to write the region of (\ref{innerwho}) as
\begin{align}
\overline{\text{co}} \Bigg[ \bigcup_{p(x_1)} \bigg\{ (R_1,R_2):
R_1 & \leq H(Y_1|X_2=\tilde{x}_2)-\eta \nonumber \\
R_2 & \leq \tau-H(Y_2|X_2=\tilde{x}_2) \bigg\} \Bigg]
\label{inner4}
\end{align}
where $\tilde{x}_2$ is given in condition 5.

Let us define $F(c)$ as
\begin{align}
F(c)=\min_{\begin{array}{c}p(x_1) \\
H(Y_1|X_2=\tilde{x}_2) = c
\end{array}} H(Y_2|X_2=\tilde{x}_2)
\label{defineF}
\end{align}
where the entropies are calculated according to the distribution
\begin{align}
p(y_1,y_2,x_1|\tilde{x}_2)=p(x_1)p(y_1|x_1,\tilde{x}_2)p'(y_2|y_1)
\end{align}
In (\ref{defineF}), we can write $\min$ instead of $\inf$ by
the same reasoning as in \cite[Section I]{Witsenhausen:1974}. Note
that $F(c)$ is not a function of $\tilde{x}_2$ because of
(\ref{tildex2}) and (\ref{usecon2}). Again, by condition 3, we can
show that $\eta \leq c \leq \log |\mathcal{Y}_1|$. Hence, the
achievable region in (\ref{inner4}) can be written as,
\begin{align}
\overline{\text{co}} \Bigg[ \bigcup_{\eta \leq c \leq \log
|\mathcal{Y}_1|} \bigg\{ (R_1,R_2): R_1 & \leq c-\eta \nonumber \\
R_2 & \leq \tau-F(c) \bigg\} \Bigg]
\end{align}
which by \cite[Facts 4 and 5]{Benzel:1979}, can further be written
as
\begin{align}
\bigcup_{\eta \leq c \leq \log |\mathcal{Y}_1|} \bigg\{ (R_1,R_2):
R_1 & \leq c-\eta \nonumber \\
 R_2 & \leq \tau- \underline{\text{env}}F(c)
\bigg\} \label{innerreal}
\end{align}
where $\underline{\text{env}}F(\cdot)$ denotes the lower convex
envelope of the function $F(\cdot)$.

\section{The Capacity Region}
In this section, we show that the achievable region in
(\ref{innerreal}) contains the outer bound in (\ref{outerreal}),
and thus, (\ref{outerreal}) and (\ref{innerreal}) are both, in
fact, single-letter characterizations of the capacity region of
DDICs satisfying conditions 1-5. To show this, it suffices to
prove that
\begin{align}
T(c) \leq \tau- \underline{\text{env}} F(c), \qquad \eta \leq c
\leq \log |\mathcal{Y}_1|
\end{align}

Let us fix a $c \in \left[\eta,\log |\mathcal{Y}_1|\right]$. Let
$p^*(u), p^*(x_1,x_2|u)$ be the distributions that achieve the
maximum in (\ref{defineT}), i.e.,
\begin{align}
H(Y_1|U)&=c \\
 I(U;Y_2) &= T(c)
\end{align}
Using condition 5, for each $u \in \mathcal{U}$, there exists a
$p^u(x_1)=\pv^u$ and a permutation matrix $G^u \in \mathcal{G}$,
such that
\begin{align}
\sum_{x_1,x_2} p^*(x_1,x_2|U=u) \pv_{x_1,x_2} = G^u
T_{\tilde{x}_2} \pv^u \label{usecon4}
\end{align}
Thus, we have
\begin{align}
H(Y_1|U=u)=H\left(G^u T_{\tilde{x}_2} \pv^u \right)=H
\left(T_{\tilde{x}_2} \pv^u \right) \label{later1}
\end{align}
(\ref{later1}) means that $\pv^u$ is in the feasible set of the
optimization in (\ref{defineF}) when $c=H(Y_1|U=u)$. Hence,
\begin{align}
F\left(H\left(Y_1|U=u\right)\right) &\leq H
\left(T'T_{\tilde{x}_2}\pv^u \right) \label{later4}
\end{align}
We have
\begin{align}
H(Y_2|U=u) &= H \left(T' G^u T_{\tilde{x}_2} \pv^u
\right) \label{IEEE3} \\
&= H \left(\Pi^u T'  T_{\tilde{x}_2} \pv^u
\right) \label{IEEE4} \\
&=H \left(T'  T_{\tilde{x}_2} \pv^u  \right)\\
&\geq F\left(H\left(Y_1|U=u\right)\right) \label{later3}
\end{align}
where
(\ref{IEEE3}), (\ref{IEEE4}) and (\ref{later3}) follow from
(\ref{usecon4}), the fact that $G \in \mathcal{G}$, and (\ref{later4}), respectively.
Thus,
\begin{align}
H(Y_2|U)&=\sum_{u}P(U=u) H(Y_2|U=u) \\
& \geq \sum_{u}P(U=u) F\left(H\left(Y_1|U=u\right)\right) \label{later6}\\
& \geq  \sum_{u}P(U=u) \underline{\text{env}}F\left(H\left(Y_1|U=u\right)\right) \label{later7}\\
& \geq \underline{\text{env}}F\left(\sum_{u}P(U=u)H\left(Y_1|U=u\right)\right) \label{later8}\\
& = \underline{\text{env}}F\left(H\left(Y_1|U\right)\right)\\
& = \underline{\text{env}}F(c) \label{later10}
\end{align}
where (\ref{later6}) follows from (\ref{later3}), (\ref{later7})
follows from the definition of $\underline{\text{env}}$, and
(\ref{later8}) follows from convexity of
$\underline{\text{env}}F(\cdot)$.

Finally, for $\eta \leq c \leq \log |\mathcal{Y}_1|$, we have
\begin{align}
T(c)&=I(U;Y_2) \\
&=H(Y_2)-H(Y_2|U) \\
&\leq \tau-\underline{\text{env}}F(c) \label{later9}
\end{align}
where (\ref{later9}) follows from (\ref{later10}) and the
definition of $\tau$ in (\ref{later11}).

Therefore, we conclude that the single-letter characterization of
the capacity region of DDICs satisfying conditions 1-5 is
(\ref{innerreal}), and also (\ref{outerreal}). To achieve point
$(R_1,R_2)$ on the boundary of the capacity region, if $R_1$ and
$R_2$ are such that
\begin{align}
R_1=c-\eta, \quad R_2 = \tau-F(c) \label{convexconvex}
\end{align}
for some $\eta \leq c \leq \log |\mathcal{Y}_1|$, transmitters 1
and 2 generate random codebooks according to $p^*(x_1)$, which is
the minimizer of $F(R_1+\eta)$, and $p^*(x_2)$, which is the
uniform distribution, respectively, and transmit the codewords
corresponding to the realizations of their own messages. Receiver
1 performs successive decoding, in the order of message 2, and
then message 1. Receiver 2 decodes its own message treating
interference from transmitter 1 as pure noise. To achieve point
$(R_1,R_2)$ on the capacity region, where $R_1$ and $R_2$ do not
satisfy (\ref{convexconvex}), time-sharing should be used. Furthermore,
we note that for these DDICs, encoder cooperation cannot increase
the capacity region.

\section{Examples}
In this section, we will provide three examples of DDICs for which
conditions 1-5 are satisfied. The first example is the channel
model adopted in \cite{Benzel:1979}, for which the capacity region
is already known. In the second and third examples, the capacity
regions are previously unknown, and using the results of this
paper, we are able to determine the capacity regions.
\subsection{Example 1} \label{first}
A DADIC is defined as \cite{Benzel:1979}
\begin{align}
Y_1&=X_1 \oplus X_2 \oplus V_1 \label{ben}\\
 Y_2&=X_1 \oplus X_2 \oplus V_1
\oplus V_2 \label{binary}
\end{align}
where
\begin{align}
\mathcal{X}_1=\mathcal{X}_2=\mathcal{Y}_1=\mathcal{Y}_2=\mathcal{S}=\{0,1,
\cdots, s-1\}
\end{align}
and $\oplus$ denotes modulo-$s$ sum, and $V_1$ and $V_2$ are
independent noise random variables defined over $\mathcal{S}$ with
distributions
\begin{align}
\pv_{i}=\left(p_i(0),p_i(1),\cdots, p_i(s-1) \right), \quad i=1,2
\end{align}
Since $Y_2=Y_1 \oplus V_2$, matrix $T'$ is circulant, and thus
input symmetric \cite[Section II.D]{Witsenhausen:1975}. Hence,
condition 1 is satisfied. It is straightforward to check that
conditions 2-5 are also satisfied. For example, when $s=3$, we
have
\begin{align}
T'&=\begin{bmatrix}  p_2(0) & p_2(2) & p_2(1) \\ p_2(1) & p_2(0) &
p_2(2) \\ p_2(2)& p_2(1) & p_2(0) \end{bmatrix}
\end{align}
and the input symmetry group for $T'$ is
\begin{align}
\mathcal{G}=\Bigg\{G_0 =\begin{bmatrix}1 & 0 & 0 \\0 & 1 & 0\\0 &
0 & 1 \end{bmatrix},\quad  G_1&=\begin{bmatrix}0 & 0 & 1 \\1 & 0 & 0\\0
& 1
& 0 \end{bmatrix}, \nonumber \\
G_2&=\begin{bmatrix}0 & 1 & 0\\ 0 & 0 & 1 \\1 & 0 &
0\end{bmatrix}\Bigg\} \label{defineG}
\end{align}
which is transitive, i.e., $ 1 \overset{G_2}{\longrightarrow} 2, 1
\overset{G_1}{\longrightarrow} 3, 2 \overset{G_1}{\longrightarrow}
1, 2 \overset{G_2}{\longrightarrow} 3, 3
\overset{G_2}{\longrightarrow} 1, 3 \overset{G_1}{\longrightarrow}
2 \nonumber $. From (\ref{ben}), we write
\begin{align}
T_{0}&=\begin{bmatrix}  p_1(0) & p_1(2) & p_1(1) \\ p_1(1) &
p_1(0)
& p_1(2) \\ p_1(2)& p_1(1) & p_1(0) \end{bmatrix} \\
T_{1}&=\begin{bmatrix}p_1(2)& p_1(1) & p_1(0) \\  p_1(0) & p_1(2)
& p_1(1) \\ p_1(1) & p_1(0)
& p_1(2)  \end{bmatrix}\\
 T_{2}  &= \begin{bmatrix}  p_1(1) &
p_1(0) & p_1(2) \\ p_1(2)& p_1(1) & p_1(0) \\ p_1(0) & p_1(2) &
p_1(1)  \end{bmatrix}
\end{align}
Conditions 2-4 are satisfied because
\begin{align}
T_1=G_1 T_0 &, \quad T_2=G_2 T_0\\
\eta&=H(V_1) \\
\sum_{x_2} p(y_1|x_1,x_2)&=p_1(0)+p_1(1)+p_1(2)=1
\end{align}
Next, we check condition 5.
\begin{align}
&\bigg\{\sum_{x_1,x_2} a_{x_1,x_2} \pv_{x_1,x_2}: \sum_{x_1,x_2}
a_{x_1,x_2}=1, a_{x_1,x_2} \geq 0 \bigg\} \label{set}\\
&= \Bigg\{a\begin{pmatrix} p_1(0) \\ p_1(1) \\p_1(2)\end{pmatrix}
+b\begin{pmatrix} p_1(2)\\p_1(0) \\ p_1(1) \end{pmatrix}
+c\begin{pmatrix}  p_1(1)\\ p_1(2)\\p_1(0)  \end{pmatrix}:\nonumber \\
&\hspace{0.3in} a+b+c=1, a,b,c \geq 0\Bigg\}
\end{align}
because even though (\ref{set}) is a convex combination of $9$
vectors, due to vectors repeating themselves in the columns of
$T_0$, $T_1$ and $T_2$, the set, in fact, consists of convex
combinations of only $3$ vectors. On the other hand, for
$\tilde{x}_2=0$,
\begin{align}
& \bigg\{G \bigg( \sum_{x_1} b_{x_1}
\pv_{x_1,\tilde{x}_2}\bigg): \sum_{x_1} b_{x_1}=1, b_{x_1} \geq 0, G =G_0 \bigg\} \label{set2}\\
&=\Bigg\{a\begin{pmatrix} p_1(0) \\ p_1(1) \\p_1(2)\end{pmatrix}
+b\begin{pmatrix} p_1(2)\\p_1(0) \\ p_1(1) \end{pmatrix}
+c\begin{pmatrix}  p_1(1)\\ p_1(2)\\p_1(0)  \end{pmatrix}:\nonumber \\
&\hspace{0.3in} a+b+c=1, a,b,c \geq 0\Bigg\}
\end{align}
because (\ref{set2}) is the convex combinations of the columns of
$T_0$, with the unitary permutation. Thus,
\begin{align}
&\bigg\{\sum_{x_1,x_2} a_{x_1,x_2} \pv_{x_1,x_2}: \sum_{x_1,x_2}
a_{x_1,x_2}=1, a_{x_1,x_2} \geq 0 \bigg\} \nonumber \\
& = \bigg\{G \bigg( \sum_{x_1} b_{x_1}
\pv_{x_1,\tilde{x}_2}\bigg): \sum_{x_1} b_{x_1}=1, b_{x_1} \geq 0, G =G_0 \bigg\}\\
& \subseteq \bigg\{G \bigg( \sum_{x_1} b_{x_1}
\pv_{x_1,\tilde{x}_2}\bigg): \sum_{x_1} b_{x_1}=1,  b_{x_1} \geq
0, G \in \mathcal{G} \bigg\}
\end{align}
and condition 5 is satisfied.

\subsection{Example 2}
Next, we consider the following DDIC. We have $
|\mathcal{X}_1|=|\mathcal{X}_2|=|\mathcal{Y}_1|=2,
|\mathcal{Y}_2|=3 $, and $p(y_1|x_1,x_2)$ is characterized by
\begin{align}
Y_1=X_1 \oplus X_2 \oplus V_1
\end{align}
where $V_1$ is Bernoulli with $p$. $p'(y_2|y_1)$ is an erasure
channel with parameter $0 \leq \alpha \leq 1$, i.e., the
transition probability matrix is
\begin{align}
T'=\begin{bmatrix} 1-\alpha & 0\\
\alpha & \alpha \\
0 & 1-\alpha \end{bmatrix}
\end{align}
Thus, the channel is such that the ``bad'' receiver cannot receive
all the bits that the ``good'' receiver receives. More
specifically, $\alpha$ proportion of the time, whether the bit is
a 0 or 1 is unrecognizable, and thus denoted as an erasure $e$.

It is easy to see that $T'$ is input symmetric because the input
symmetry group
\begin{align}
\mathcal{G}=\Bigg\{\begin{bmatrix} 1 & 0 \\ 0 &1 \end{bmatrix},
\begin{bmatrix} 0 & 1 \\ 1 &0 \end{bmatrix} \Bigg\}
\end{align}
is transitive. Conditions 2-5 are satisfied because
$p(y_1|x_1,x_2)$ is the same as in Example 1 in Section
\ref{first}.

\subsection{Example 3}
\begin{figure}
\centering
\includegraphics[width=3.5in]{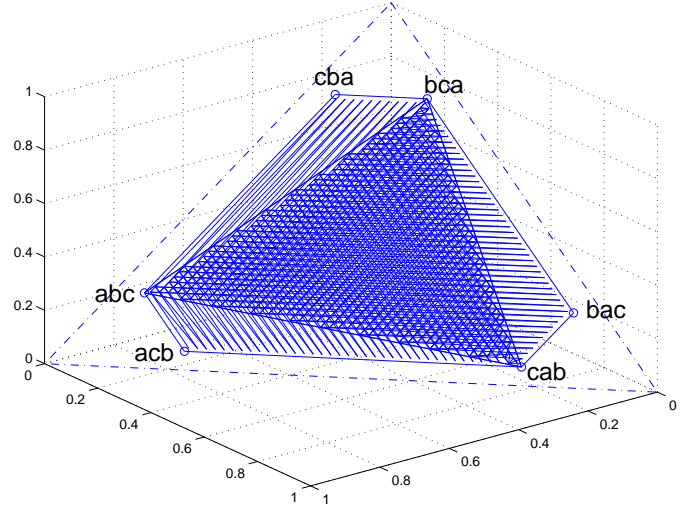}
\caption{Explanation of condition 5 in example 3.} \label{eg3fig}
\end{figure}

Let $a,b,c, d,e,f$ be non-negative numbers such that $a+b+c=1$ and
$d+e+f=1/2$. We have $|\mathcal{X}_1|=4$,
$|\mathcal{X}_2|=|\mathcal{Y}_1|=3$, and $|\mathcal{Y}_2|=6$. The
DDIC is described as
\begin{align}
T'&=\begin{bmatrix}d & e & f\\
e& f&d\\
d & f& e\\
f& e& d\\
e& d& f\\
f& d& e \end{bmatrix}\\
T_{0} &= \begin{bmatrix}
a&b&c&c\\
b&c&a&b\\
c&a&b&a
\end{bmatrix} \\
T_{1} &=\begin{bmatrix}
c&a&b&a\\
a&b&c&c\\
b&c&a&b
\end{bmatrix}  \\
T_2 &=\begin{bmatrix}
b&c&a&b\\
c&a&b&a\\
a&b&c&c
\end{bmatrix}
\end{align}
It is straightforward to see that $T'$ is input symmetric because
the input symmetry group
\begin{align}
\mathcal{G}=\Bigg\{G_0 &=\begin{bmatrix}1 & 0 & 0 \\0 & 1 & 0\\0 &
0 & 1 \end{bmatrix}, G_1=\begin{bmatrix}0 & 0 & 1 \\1 & 0 & 0\\0 &
1
& 0 \end{bmatrix}, \nonumber \\
G_2&=\begin{bmatrix}0 & 1 & 0\\ 0 & 0 & 1 \\1 & 0 &
0\end{bmatrix}, G_3=\begin{bmatrix}1 & 0 & 0\\ 0 & 0 & 1 \\0 & 1 &
0\end{bmatrix}, \nonumber \\
G_4&=\begin{bmatrix}0 & 1 & 0\\ 1 & 0 & 0 \\0 & 0 &
1\end{bmatrix}, G_5=\begin{bmatrix}0 & 0 & 1\\ 0 & 1 & 0 \\1 & 0 &
0\end{bmatrix}\Bigg\}
\end{align}
is transitive. Conditions 2-4 are satisfied because
\begin{align}
T_1  = G_1 T_0 &, \quad
T_2  = G_2 T_0 \\
\eta=-a \log a -& b \log b-c \log c \\
\sum_{x_2} p(y_1|x_1,x_2)&=a+b+c=1
\end{align}
To show condition 5, we use Figure \ref{eg3fig}. The set on the
first line of (\ref{condition42}) in condition 5 is the convex
combination of the following six points,
\begin{align}
\begin{bmatrix} a \\ b \\ c \end{bmatrix},
\begin{bmatrix} a \\ c \\ b \end{bmatrix},
\begin{bmatrix} c \\ a \\ b \end{bmatrix},
\begin{bmatrix} b \\ a \\ c \end{bmatrix},
\begin{bmatrix} b \\ c \\ a \end{bmatrix},
\begin{bmatrix} c \\ b \\ a \end{bmatrix}
\end{align}
resulting in all the points within the hexagon in Figure
\ref{eg3fig}. The three sets
\begin{align}
&\bigg\{G \bigg( \sum_{x_1} b_{x_1}
\pv_{x_1,\tilde{x}_2}\bigg): \sum_{x_1} b_{x_1}=1, b_{x_1} \geq 0, G=G_0 \bigg\} \nonumber \\
=&\Bigg\{\mu_1\begin{bmatrix} a \\ b \\ c
\end{bmatrix}+\mu_2\begin{bmatrix} b \\ c \\ a \end{bmatrix}
+\mu_3\begin{bmatrix} c \\a \\ b \end{bmatrix}+\mu_4\begin{bmatrix} c \\ b \\ a \end{bmatrix}: \nonumber \\
&\hspace{1.5in}\sum_{i=1}^4 \mu_i=1,\mu_i \geq 0 \Bigg\}
\end{align}
and
\begin{align}
&\bigg\{G \bigg( \sum_{x_1} b_{x_1}
\pv_{x_1,\tilde{x}_2}\bigg): \sum_{x_1} b_{x_1}=1, b_{x_1} \geq 0, G=G_1 \bigg\} \nonumber \\
=&\Bigg\{\mu_1\begin{bmatrix} c \\ a \\ b
\end{bmatrix}+\mu_2\begin{bmatrix} a \\ b \\ c \end{bmatrix}
+\mu_3\begin{bmatrix} b \\ c \\ a \end{bmatrix}+\mu_4\begin{bmatrix} a \\ c \\ b \end{bmatrix}: \nonumber \\
&\hspace{1.5in}\sum_{i=1}^4 \mu_i=1,\mu_i \geq 0 \Bigg\}
\end{align}
and
\begin{align}
&\bigg\{G \bigg( \sum_{x_1} b_{x_1}
\pv_{x_1,\tilde{x}_2}\bigg): \sum_{x_1} b_{x_1}=1, b_{x_1} \geq 0, G =G_2 \bigg\} \nonumber \\
=&\Bigg\{\mu_1\begin{bmatrix} b \\ c \\ a
\end{bmatrix}+\mu_2\begin{bmatrix} c \\ a \\ b \end{bmatrix}
+\mu_3\begin{bmatrix} a \\ b \\ c \end{bmatrix}+\mu_4\begin{bmatrix} b \\ a \\ c \end{bmatrix}: \nonumber \\
&\hspace{1.5in}\sum_{i=1}^4 \mu_i=1,\mu_i \geq 0 \Bigg\}
\end{align}
correspond to the points in the three shaded areas, [$abc$, $cba$,
$bca$, $cab$], $[acb,abc,bca,cab]$, and $[bac, cab, abc, bca]$,
respectively. Since the three shaded areas cover the entire hexagon, and
$\{G_0,G_1,G_2\} \subset \mathcal{G}$, condition 5 is satisfied.

\section{Conclusion}
We provide a single-letter characterization for the capacity
region of a class of DDICs, which is more general than the class
of DADICs studied by Benzel\cite{Benzel:1979}. We show that for the
class of DDICs studied, encoder cooperation does not increase the
capacity region, and the best way to manage the interference is
through random codebook design and treating the signal for the
``good'' receiver as pure noise at the ``bad'' receiver.

\bibliographystyle{IEEEtran}
\bibliography{IEEEabrv,ref}

\end{document}